\documentclass
[aps,prd,onecolumn,superscriptaddress,preprintnumbers,nofootinbib]{revtex4}%
\usepackage{amsfonts}
\usepackage{amsmath}
\usepackage{graphicx}
\usepackage{wrapfig}
\usepackage[hypertexnames=false]{hyperref}
\usepackage{enumerate}%
\usepackage{amssymb}
\providecommand{\U}[1]{\protect\rule{.1in}{.1in}}
\tighten  
\begin{document}
\title{Asymptotically flat spacetimes in three-dimensional higher spin gravity}
\author{Hern\'an A. Gonz\'alez}
\email{hdgonzal-at-uc.cl}
\affiliation{Universidad Andr\'es Bello, Av. Rep\'ublica 440, Santiago, Chile.}
\affiliation{Facultad de F\'isica, Pontificia Universidad Cat\'{o}lica de Chile, Casilla
306, Santiago 22, Chile.}
\author{Javier Matulich}
\email{matulich-at-cecs.cl}
\affiliation{Centro de Estudios Cient\'ificos (CECs), Casilla 1469, Valdivia, Chile.}
\affiliation{Departamento de F\'isica, Universidad de Concepci\'on, Casilla, 160-C,
Concepci\'on, Chile.}
\author{Miguel Pino}
\email{miguel.pino.r-at-usach.cl}
\affiliation{Departamento de F\'isica, Universidad de Santiago de Chile, Av. Ecuador 3493,
Estaci\'on Central, Santiago, Chile.}
\author{Ricardo Troncoso}
\email{troncoso-at-cecs.cl}
\affiliation{Centro de Estudios Cient\'ificos (CECs), Casilla 1469, Valdivia, Chile.}
\affiliation{Universidad Andr\'es Bello, Av. Rep\'ublica 440, Santiago, Chile.}
\preprint{CECS-PHY-13/06}

\begin{abstract}
A consistent set of asymptotic conditions for higher spin gravity in three
dimensions is proposed in the case of vanishing cosmological constant. The
asymptotic symmetries are found to be spanned by a higher spin extension of
the BMS$_{3}$ algebra with an appropriate central extension. It is also shown
that our results can be recovered from the ones recently found for
asymptotically AdS$_{3}$ spacetimes by virtue of a suitable gauge choice that
allows to perform the vanishing cosmological constant limit.

\end{abstract}
\maketitle

\section{Introduction}

Higher spin gravity \cite{VV0,VV1,VV2} has recently motivated a flurry of
activity. In the case of $d\geq4$ dimensions, the consistency of the theory
necessarily requires the presence of a negative cosmological constant, which
naturally makes it to be appealing in the context of holographic dualities
\cite{KlePo, Gaberdiel-Gopakumar} (For recent reviews see e.g.,
\cite{ReviewHS0,ReviewHS1,ReviewHS2,ReviewHS3,ReviewHS4, ReviewHS5}). The
three-dimensional case becomes particularly interesting in order to acquire a
deeper understanding of the subject. Indeed, in this case the generic theory
turns out to be exceptionally simpler, since it can be described as a standard
field theory in terms of a Chern-Simons action. Moreover, it can be further
simplified because it admits a consistent truncation that characterizes the
dynamics of a finite number of higher spin gauge fields \cite{Blencowe,BBS}.
Remarkably, the theory in $d=3$ can also be formulated in the case of
vanishing cosmological constant $\Lambda$. Therefore, this latter specific
case naturally becomes worth of special attention, since one may naturally
expect that it could unveil some clues about the elusive construction of
higher spin gravity around flat spacetime in higher dimensions.

The purpose of this work is to propose a precise set of asymptotic conditions
for higher spin gravity in three dimensions in the case of $\Lambda=0$. In the
next section, the theory is revisited in some detail, for simplicity, in the
case of spins $s=2$, $3$. Section \ref{vanishing lambda} is devoted to present
the asymptotically flat conditions, including the analysis of their asymptotic
symmetry group. The corresponding conserved charges are also shown to span a
higher spin extension of the BMS$_{3}$ algebra with a nontrivial central
extension. In section \ref{Limit}, we show that these results can be recovered
from the corresponding ones in the case of asymptotically AdS$_{3}$ spacetimes
\cite{Henneaux-HS,Theisen-HS}. This task is performed through a particular
\textquotedblleft decoupling\textquotedblright\ gauge choice, that allows the
straightforward application of the vanishing $\Lambda$ limit. The procedure
allows extending the asymptotically flat conditions to the case of spins
$s\geq2$. Finally, the discussion is carried out in section \ref{Discussion}%
.\bigskip

Note added: While this work was being typed, ref. \cite{ABFGR} was posted on
the arXiv, whose results overlap with some of ours in sections
\ref{vanishing lambda}, \ref{Limit}.

\section{Higher spin gravity in 3D with vanishing cosmological constant}

Higher spin gravity in three dimensions can be formulated in terms of a
Chern-Simons action \cite{Blencowe,BBS}, given by,
\begin{equation}
I[A]=\frac{k}{4\pi}\int\langle AdA+\frac{2}{3}A^{3}\rangle\ .\label{CS2}%
\end{equation}
For the sake of simplicity, in this section we discuss in some detail the case
of spins $s=2$, $3$. In this case, the gauge field $A=A_{\mu}dx^{\mu}$ can be
written as (see e.g., \cite{Henneaux-HS,Theisen-HS})
\begin{equation}
A=\omega^{a}J_{a}+e^{a}P_{a}+W^{ab}J_{ab}+E^{ab}P_{ab}\ ,
\end{equation}
where the set $\left\{  J_{a},P_{a},J_{ab},P_{ab}\right\}  $\ spans the gauge
group, and the generators $P_{ab}$, $J_{ab}$ are assumed to be symmetric and
traceless. As explained in \cite{Theisen-HS}, in the case of vanishing
cosmological constant, the generators fulfill a generalization of the
Poincar\'{e} algebra, given by\footnote{This algebra can be obtained from a
contraction of two copies of $sl(3)$ (see appendix \ref{contraction}).}
\begin{equation}%
\begin{split}
&  [J_{a},J_{b}]=\epsilon_{abc}J^{c},\quad{}[P_{a},J_{b}]=\epsilon_{abc}%
P^{c},\quad{}[P_{a},P_{b}]=0,\\
& \\
&  {}[J_{a},J_{bc}]=\epsilon_{\;\;a(b}^{m}J_{c)m},\quad\lbrack J_{a}%
,P_{bc}]=\epsilon_{\;\;a(b}^{m}P_{c)m},\\
&  {}[P_{a},J_{bc}]=\epsilon_{\;\;a(b}^{m}P_{c)m},\quad{}[P_{a},P_{bc}]=0,\\
& \\
&  {}[J_{ab},J_{cd}]=-\left(  \eta_{a(c}\epsilon_{d)bm}+\eta_{b(c}%
\epsilon_{d)am}\right)  J^{m},\quad{}[J_{ab},P_{cd}]=-\left(  \eta
_{a(c}\epsilon_{d)bm}+\eta_{b(c}\epsilon_{d)am}\right)  P^{m},\\
&  {}[P_{ab},J_{cd}]=-\left(  \eta_{a(c}\epsilon_{d)bm}+\eta_{b(c}%
\epsilon_{d)am}\right)  P^{m},\quad{}[P_{ab},P_{cd}]=0,
\end{split}
\label{CS1}%
\end{equation}
the level is related to the Newton constant according to $k=\frac{1}{4G}$, and
the bracket $\langle\cdot\cdot\cdot\rangle$ in (\ref{CS2}) stands for a
non-degenerate invariant bilinear product, whose only nonvanishing components
are given by\footnote{This is related to the fact that the algebra (\ref{CS1})
admits a Casimir operator given by $C=P^{a}J_{a}+\frac{1}{2}P^{ab}J_{ab}$.}%
\begin{equation}
\langle P_{a}J_{b}\rangle=\eta_{ab},\quad\langle P_{ab}J_{cd}\rangle=\eta
_{ac}\eta_{bd}+\eta_{ad}\eta_{cb}-\frac{2}{3}\eta_{ab}\eta_{cd}\ .\label{CS3}%
\end{equation}
Therefore, the action (\ref{CS2}) reduces to
\begin{equation}
I[e,\omega,E,W]=\frac{k}{2\pi}\int\left[  e^{a}\left(  d\omega_{a}+\frac{1}%
{2}\epsilon_{abc}\omega^{b}\omega^{c}+2\epsilon_{abc}W^{bd}W_{\;\;d}%
^{c}\right)  +2E^{ab}(dW_{ab}+2\epsilon_{cda}\omega^{c}W_{b}^{\;d})\right]
\ ,\label{CS7}%
\end{equation}
up to a boundary term, and the field equations read
\begin{subequations}
\label{CS8}%
\begin{align}
de^{a}+\epsilon^{abc}\omega_{b}e_{c}+4\epsilon^{abc}E_{bd}W_{\;\;c}^{d}  &
=0,\\
d\omega^{a}+\frac{1}{2}\epsilon^{abc}\omega_{b}\omega_{c}+2\epsilon
^{abc}W_{bd}W_{\;\;c}^{d}  & =0,\\
dE^{ab}+\epsilon^{cd(a|}\omega_{c}E_{d}^{\;\;|b)}+\epsilon^{cd(a|}e_{c}%
W_{d}^{\;\;|b)}  & =0,\\
dW^{ab}+\epsilon^{cd(a|}\omega_{c}W_{d}^{\;\;|b)}  & =0.
\end{align}
The fields $e^{a}$, $E^{ab}$, and $\omega^{a}$, $W^{ab}$ are interpreted as a
generalization of the dreibein and the spin connection, respectively; so that
the metric and the spin-3 field can be constructed from the only quadratic and
cubic combinations of the generalized dreibein that are invariant under
Lorentz-like transformations generated by $J_{a}$, $J_{ab}$, i.e.
\end{subequations}
\begin{align}
ds^{2}  & =(\eta_{ab}e_{\mu}^{a}e_{\nu}^{b}+2\eta_{ac}\eta_{bd}E_{\mu}%
^{ab}E_{\nu}^{cd})dx^{\mu}dx^{\nu},\label{metric}\\
\varphi & =(\eta_{ac}\eta_{bd}e_{\mu}^{a}e_{\nu}^{b}E_{\rho}^{cd}-\frac{4}%
{3}\eta_{cd}\eta_{ag}\eta_{bh}E_{\mu}^{gc}E_{\nu}^{hd}E_{\rho}^{ab})dx^{\mu
}dx^{\nu}dx^{\rho}.\label{spin3 field}%
\end{align}
Note that if $e_{\mu}^{a}$ is assumed to be invertible, then in the case of
$E^{ab}=0$, the field equations imply that $W^{ab}$ also vanishes, and hence,
General Relativity in vacuum is recovered.

\section{Asymptotically flat spacetimes endowed with higher spin fields}

\label{vanishing lambda}

It has been shown that asymptotically flat spacetimes in General Relativity
with vanishing cosmological constant in three dimensions enjoy similar
features as the ones exhibited by asymptotically AdS$_{3}$ geometries
\cite{Brown-Henneaux}. Indeed, the asymptotic symmetry group is infinite
dimensional \cite{Ashtekar}, and its algebra, so-called BMS$_{3}$, also
acquires a nontrivial central extension \cite{Barnich-Compere,
Barnich-Troessaert}. It has also been recently shown that these results can be
recovered from a suitable Penrose-like limit of the asymptotic behaviour of
the metric on AdS$_{3}$ \cite{Limite-plano}. The BMS$_{3}$ algebra turns out
to be isomorphic to the Galilean conformal algebra in two dimensions
(GCA$_{2}$) \cite{BagchiGCA}, which becomes relevant in the context of
non-relativistic holography.

It is then natural to look for suitable asymptotic conditions in the case of
higher spin gravity, being such that they fulfill the following requirements:

\begin{enumerate}
[(i)]

\item Reducing to the ones of \cite{Barnich-Compere,Barnich-Troessaert} when
the higher spin fields are switched off, and

\item In presence of higher spin fields, they should correspond to the
vanishing cosmological constant limit of the asymptotically AdS$_{3}$
conditions of \cite{Henneaux-HS, Theisen-HS,Campoleoni-HS,Gaberdiel-Hartman}.
\end{enumerate}

\bigskip

In this section the analysis is explicitly carried out in the simplest case of
spins $s=2$, $3$; since the generic case of higher spins is shown to be one of
the outputs of section \ref{Limit}.

For the theory under consideration, gauge fields satisfying the previous
requirements, are proposed to possess the following asymptotic
form\footnote{Hereafter, our conventions are such that we assume a
non-diagonal Minkowski metric in tangent space, whose only nonvanishing
components are given by $\eta_{01}=\eta_{10}=\eta_{22}=1$, and the Levi-Civita
symbol fulfills $\epsilon_{012}=1$.}:
\begin{multline}
A=\left(  \frac{1}{2}\mathcal{M}du-dr+\left(  \mathcal{J}+\frac{u}{2}%
\partial_{\phi}\mathcal{M}\right)  d\phi\right)  P_{0}+duP_{1}+rd\phi
P_{2}+\frac{1}{2}\mathcal{M}d\phi J_{0}+d\phi J_{1}\label{arriba}\\
+\left(  \mathcal{W}du+\left(  \mathcal{V}+u\partial_{\phi}\mathcal{W}\right)
d\phi\right)  P_{00}+\mathcal{W}d\phi J_{00},
\end{multline}
where $r,\phi$ correspond to the radial and angular coordinates, respectively,
$u$ is a null coordinate that plays the role of time, and $\mathcal{M}$,
$\mathcal{J}$, $\mathcal{W}$ and $\mathcal{V}$ stand for arbitrary functions
of $\phi$. According to (\ref{arriba}), it is apparent that condition (i) is
fulfilled in the case of $\mathcal{W}=\mathcal{V}=0$; while condition (ii) is
explicitly shown to be satisfied in the next section.

The asymptotic symmetries correspond to gauge transformations generated by a
Lie-algebra-valued parameter
\begin{equation}
\lambda=\rho^{a}P_{a}+\eta^{a}J_{a}+\xi^{ab}P_{ab}+\Lambda^{ab}J_{ab}%
\ ,\label{lambda-flat}%
\end{equation}
that preserves the form of (\ref{arriba}), i.e.,
\[
\delta A=d\lambda+[A,\lambda]=\mathcal{O}(A).
\]
One then finds that $\lambda=\lambda(\epsilon,y,w,v)$ depends on four
independent functions of the angular coordinate, being defined as%
\begin{equation}
\rho^{1}=\epsilon+uy^{\prime},\quad\eta^{1}=y,\quad\xi^{11}=w+uv^{\prime
},\quad\Lambda^{11}=v,
\end{equation}
where prime denotes the derivative with respect to $\phi$. The remaining
components of $\lambda$ are given in appendix \ref{appendix lambda}.

The arbitrary functions appearing in the asymptotic form of the gauge field
are found to transform according to the following rules
\begin{equation}%
\begin{split}
\delta\mathcal{M} &  =y\mathcal{M}^{\prime}+2y^{\prime}\mathcal{M}%
-2y^{\prime\prime\prime}+4\left(  2v\mathcal{W}^{\prime}+3v^{\prime
}\mathcal{W}\right)  ,\\
\delta\mathcal{J} &  =y\mathcal{J}^{\prime}+2y^{\prime}\mathcal{J}+\frac{1}%
{2}\epsilon\mathcal{M}^{\prime}+\epsilon^{\prime}\mathcal{M}-\epsilon
^{\prime\prime\prime}+2\left(  2w\mathcal{W}^{\prime}+3w^{\prime}%
\mathcal{W}+2v\mathcal{V}^{\prime}+3v^{\prime}\mathcal{V}\right)  ,\\
\delta\mathcal{W} &  =y\mathcal{W}^{\prime}+3y^{\prime}\mathcal{W}-\frac{1}%
{6}v\mathcal{M}^{\prime\prime\prime}-\frac{3}{4}v^{\prime}\mathcal{M}%
^{\prime\prime}-\frac{5}{4}v^{\prime\prime}\mathcal{M}-\frac{5}{6}%
v^{\prime\prime\prime}\mathcal{M}+\frac{2}{3}(v\mathcal{M}\mathcal{M}^{\prime
}+v^{\prime}\mathcal{M}^{2})+\frac{1}{6}v^{(5)},\\
\delta\mathcal{V} &  =y\mathcal{V}^{\prime}+3y^{\prime}\mathcal{V}%
+\epsilon\mathcal{W}^{\prime}+3\epsilon^{\prime}\mathcal{W}-\frac{5}%
{3}v^{\prime\prime\prime}\mathcal{J}-\frac{5}{2}v^{\prime\prime}%
\mathcal{J}^{\prime}-\frac{3}{2}v^{\prime}\mathcal{J}^{\prime\prime}-\frac
{1}{3}v\mathcal{J}^{\prime\prime\prime}+\frac{4}{3}v(\mathcal{J}%
\mathcal{M})^{\prime}\\
&  +\frac{8}{3}v^{\prime}\mathcal{M}\mathcal{J}+\frac{2}{3}(w\mathcal{M}%
\mathcal{M}^{\prime}+w^{\prime}\mathcal{M}^{2})-\frac{1}{6}w\mathcal{M}%
^{\prime\prime\prime}-\frac{3}{4}w^{\prime}\mathcal{M}^{\prime\prime}-\frac
{5}{4}w^{\prime\prime}\mathcal{M}^{\prime}-\frac{5}{6}w^{\prime\prime\prime
}\mathcal{M}+\frac{1}{6}w^{(5)},
\end{split}
\end{equation}
which allows to find the precise form of the asymptotic symmetry algebra. This
can be readily carried out as follows. The variation of the global charges
that correspond to the asymptotic symmetries spanned by $\lambda$, in the
canonical approach \cite{Regge-Teitelboim}, is given by
\begin{equation}
\delta Q[\lambda]=\frac{k}{2\pi}\int\langle\lambda\delta A_{\phi}\rangle
d\phi.\label{ss8}%
\end{equation}
Once the variation of the charges is evaluated on the asymptotic form of the
connection (\ref{arriba}), it is found that it becomes linear in the variation
of the fields, so that it can be integrated as%
\begin{equation}
Q[\epsilon,y,w,v]=\frac{k}{4\pi}\int\left[  \epsilon\mathcal{M}+2y\mathcal{J}%
+4(w\mathcal{W}+v\mathcal{V})\right]  d\phi.\label{ss9}%
\end{equation}
Their algebra can then be straightforwardly obtained from the variation of the
fields since, as explained in \cite{BrownHenneaux-2}, $\delta_{\lambda_{2}%
}Q[\lambda_{1}]=\{Q[\lambda_{1}],Q[\lambda_{2}]\}$. Indeed, as expected, the
asymptotic symmetries associated to $\epsilon(\phi)$ and $y(\phi)$ span the
BMS$_{3}$ algebra with the same central charge as in the case of General
Relativity in three dimensions \cite{Barnich-Compere}. In this case, the
Poisson brackets read
\begin{equation}
\{Q(\epsilon_{1},y_{1}),Q(\epsilon_{2},y_{2})\}=Q(\epsilon_{\lbrack
1,2]},y_{[1,2]})+K[\epsilon_{1},\epsilon_{2},y_{1},y_{2}],\label{ss10}%
\end{equation}
where the parameters $\epsilon_{\lbrack1,2]}$ and $y_{[1,2]}$ are given by
\begin{align}
\epsilon_{\lbrack1,2]} &  =\epsilon_{1}y_{2}^{\prime}-\epsilon_{2}%
y_{1}^{\prime}-\epsilon_{1}^{\prime}y_{2}+\epsilon_{2}^{\prime}y_{1}%
,\label{ss11}\\
y_{[1,2]} &  =y_{1}y_{2}^{\prime}-y_{1}^{\prime}y_{2},\label{ss12}%
\end{align}
and the central charge $K$ is
\begin{equation}
K[\epsilon_{1},\epsilon_{2},y_{1},y_{2}]=\frac{k}{2\pi}\int\left[
\epsilon_{1}^{\prime}y_{2}^{\prime\prime}-\epsilon_{2}^{\prime}y_{1}%
^{\prime\prime}\right]  d\phi.\label{ss13}%
\end{equation}
Expanding in Fourier modes $P_{n}=Q(\epsilon=e^{in\phi})$ and $J_{n}%
=Q(y=e^{in\phi})$, the algebra acquires the form%
\begin{subequations}
\label{ss14}%
\begin{align}
i\{P_{n},P_{m}\} &  =0,\\
i\{J_{n},J_{m}\} &  =(n-m)J_{n+m},\\
i\{J_{n},P_{m}\} &  =(n-m)P_{n+m}+kn^{3}\delta_{m+n}.
\end{align}
The brackets of the BMS$_{3}$ generators with the remaining charges associated
to $w(\phi)$ and $v(\phi)$, are given by
\end{subequations}
\begin{equation}
\{Q(\epsilon_{1},y_{1}),Q(w_{2},v_{2})\}=Q(w_{[1,2]},v_{[1,2]}),\label{ss15}%
\end{equation}
where $w_{[1,2]}$ and $v_{[1,2]}$ read
\begin{align}
w_{[1,2]} &  =\epsilon_{1}v_{2}^{\prime}-2\epsilon_{1}^{\prime}v_{2}%
+w_{2}^{\prime}y_{1}-2w_{2}y_{1}^{\prime},\\
v_{[1,2]} &  =y_{1}v_{2}^{\prime}-2v_{2}y_{1}^{\prime}.
\end{align}
Therefore, the corresponding Fourier modes, $W_{n}=Q(w=e^{in\phi})$ and
$V_{n}=Q(v=e^{in\phi})$, fulfill
\begin{subequations}
\label{ss16}%
\begin{align}
i\{P_{n},W_{m}\} &  =0,\\
i\{J_{n},W_{m}\} &  =(2n-m)W_{n+m},\\
i\{P_{n},V_{m}\} &  =(2n-m)W_{n+m},\\
i\{J_{n},V_{m}\} &  =(2n-m)V_{n+m},
\end{align}
from which, one verifies that the charges $W_{n}$ and $V_{n}$ correspond to
spin-3 generators. Finally, the brackets of the spin-3 generators turn out to
be given by
\end{subequations}
\begin{subequations}
\label{ss17}%
\begin{align}
i\{W_{n},W_{m}\} &  =0,\\
i\{W_{n},V_{m}\} &  =\frac{1}{3}\left[  \frac{8}{k}(n-m)\sum_{j=-\infty
}^{\infty}P_{j}P_{n+m-j}+(n-m)(2n^{2}+2m^{2}-mn)P_{m+n}+kn^{5}\delta
_{m+n}\right]  ,\\
i\{V_{n},V_{m}\} &  =\frac{1}{3}\left[  \frac{16}{k}(n-m)\sum_{j=-\infty
}^{\infty}P_{j}J_{n+m-j}+(n-m)(2n^{2}+2m^{2}-mn)J_{m+n}\right]  ,
\end{align}
so that the nonlinearity of the algebra becomes manifest. Note that the
structure of the brackets of the spin-3 generators is reminiscent of the one
of BMS$_{3}$, in the sense that $W_{n}$ can be regarded as a higher spin
extension of the supertranslations $P_{n}$, and the central term appears in
the crossed bracket.

In sum, the commutation relations \eqref{ss14}, \eqref{ss16} and \eqref{ss17}
provide the searched for higher spin extension of the BMS$_{3}$ algebra.

It is interesting to express the asymptotic form of the connection $A$ in
(\ref{arriba}), in terms of the metric and the spin-3 field, which read
\end{subequations}
\begin{align*}
ds^{2}  & =\mathcal{M}du^{2}-2dudr+2\left(  \mathcal{J}+\frac{u}{2}%
\partial_{\phi}\mathcal{M}\right)  dud\phi+r^{2}d\phi^{2},\\
\varphi & =\left(  \mathcal{W}du+\left(  \mathcal{V}+u\partial_{\phi
}\mathcal{W}\right)  d\phi\right)  du^{2}.
\end{align*}
respectively. From this, it can be explicitly seen that for our gauge choice,
the fields decouple, in the sense that the metric does not acquire any kind of
backreaction from the spin-3 field, neither the spin-3 field does from the
metric. Furthermore, unlike the case of $U(1)$ fields coupled to General
Relativity (see e.g., \cite{Kraus-lectures}), the spin-3 field does not
contribute to neither the energy nor the angular momentum. Hence, in spite of
these unusual features between the coupling of the spin-3 field with the
metric, it is amusing to verify that the charges still fulfill a rather
nontrivial algebra that mixes the asymptotic form of both fields in a
nonlinear form. It is also worth pointing out that a similar effect is known
to occur for a class of configurations in the case of scalar fields
non-minimally coupled to gravity \cite{Eloy-Cheshire}. The next section is
devoted to show that our asymptotic conditions fulfill requirement (ii), as
well as to provide the extension of these results to the case of spins
$s\geq2$.

\section{Vanishing cosmological constant limit of AdS$_{3}$ boundary
conditions}

\label{Limit}

Higher spin gravity on AdS$_{3}$ is described through the difference of two
Chern-Simons actions, whose gauge fields correspond to two copies of a single
algebra \cite{Blencowe, BBS}. The asymptotic behavior of the fields, in this
case, has been discussed in \cite{Henneaux-HS,
Theisen-HS,Campoleoni-HS,Gaberdiel-Hartman}, for $sl(N)$ and $hs(\lambda)$. In
the \textquotedblleft highest weight gauge\textquotedblright, the asymptotic
form of the gauge fields can be written as
\begin{equation}
A^{\pm}=b_{\pm}^{-1}a^{\pm}b_{\pm}+b_{\pm}^{-1}db_{\pm},
\end{equation}
where $b_{\pm}=e^{\pm\log(r/l)L_{0}}$, and%
\[
a^{\pm}=\pm(L_{\pm1}-\Xi_{\pm}L_{\mp1}-W_{\pm}W_{\mp2}+\cdot\cdot\cdot
)dx^{\pm}.
\]
Here $\Lambda=-\frac{1}{l^2}$, $L_{i}$ describes the generators of $sl(2)$ in the principal embedding,
$\Xi_{\pm}$ and $W_{\pm}$ correspond to arbitrary functions of $x^{\pm}%
=\frac{t}{l}\pm\phi$, and the dots stand for additional terms involving the
higher spin generators ($s>3$) of highest weight.

For our purposes, it is convenient to make a different gauge choice, so that
the asymptotic form of the gauge fields read%
\[
A^{\pm}=g_{\pm}{}^{-1}a^{\pm}g_{\pm}+g_{\pm}{}^{-1}dg_{\pm},
\]
with%
\begin{align}
g_{+} &  =b_{+}e^{-\log(\sqrt{2}\frac{r}{l})L_{0}}e^{\frac{r}{\sqrt{2}l}%
L_{-1}},\label{g+}\\
g_{-} &  =b_{-}e^{-\log(\frac{1}{2\sqrt{2}}\frac{r}{l})L_{0}}e^{\frac{r}%
{\sqrt{2}l}L_{-1}}e^{\sqrt{2}\frac{l}{r}L_{1}}.\label{g-}%
\end{align}
Therefore, the components of the connections are given by%
\begin{equation}
A^{\pm}=\frac{r}{l}dx^{\pm}L_{0}^{\pm}\pm\frac{1}{\sqrt{2}}\left[  \frac
{dr}{l}+\left(  \frac{r^{2}}{2l^{2}}-2\Xi_{\pm}\right)  dx^{\pm}\right]
L_{-1}^{\pm}\pm\frac{dx^{\pm}}{\sqrt{2}}L_{1}^{\pm}\mp2dx^{\pm}(W_{\pm}%
W_{-2}^{\pm}+\cdot\cdot\cdot).\label{A AdS our gauge}%
\end{equation}
As it is explained below, it is convenient to make the change $t=u$, since in
this gauge, $u$ becomes a null coordinate. In order to explore some of the
features of this gauge choice, it is instructive to discuss the case of
$sl(3)$ with more detail. In this case, it has been shown that the asymptotic
symmetries are generated by two copies of the $W_{3}$ algebra
\cite{Henneaux-HS}, \cite{Theisen-HS}, defined by
\begin{align}
i\{\mathcal{L}^{\pm}_{n},\mathcal{L}^{\pm}_{m}\} & =(n-m)\mathcal{L}^{\pm
}_{n+m}+\frac{kl}{2}n^{3}\delta_{m+n},\nonumber\\
i\{\mathcal{L}^{\pm}_{n},\mathcal{W}^{\pm}_{m}\} & =(2n-m)\mathcal{W}^{\pm
}_{n+m},\label{W3}\\
i\{\mathcal{W}^{\pm}_{n},\mathcal{W}^{\pm}_{m}\} & =\frac{1}{3}\left[
\frac{16}{kl}(n-m)\sum_{j=-\infty}^{\infty}\mathcal{L}^{\pm}_{j}%
\mathcal{L}^{\pm}_{n+m-j}+(n-m)(2n^{2}+2m^{2}-mn)\mathcal{L}^{\pm}_{m+n}%
+\frac{kl}{2}n^{5}\delta_{m+n}\right] .\nonumber
\end{align}

It is also useful to change the basis according to
\begin{equation}
L_{-1}^{\pm}=-\sqrt{2}J_{0}^{\pm},\quad L_{0}^{\pm}=J_{2}^{\pm},\quad
L_{1}^{\pm}=\sqrt{2}J_{1}^{\pm},
\end{equation}%
\begin{equation}
W_{-2}^{\pm}=-2T_{00}^{\pm},\quad W_{-1}^{\pm}=\sqrt{2}T_{02}^{\pm},\quad
W_{0}^{\pm}=-T_{22}^{\pm},\quad W_{1}^{\pm}=-\sqrt{2}T_{12}^{\pm},\quad
W_{2}^{\pm}=-2T_{11}^{\pm},
\end{equation}
where the generators $T_{ab}$ are traceless, followed by
\begin{equation}
J_{a}^{\pm}=\frac{J_{a}\pm lP_{a}}{2}\quad\quad\quad\quad T_{ab}^{\pm}%
=\frac{J_{ab}\pm lP_{ab}}{2},
\end{equation}
so that the full gauge field reads
\begin{align}
A=A^{+}+A^{-}= &  \left(  \frac{1}{2}\mathcal{M}d\phi+\frac{\mathcal{N}}%
{l^{2}}du-\frac{r^{2}}{2l^{2}}d\phi\right)  J_{0}+d\phi J_{1}+\frac{r}{l^{2}%
}duJ_{2}\nonumber\\
&  +\left(  -dr+\frac{1}{2}\mathcal{M}du+\mathcal{N}d\phi-\frac{r^{2}}{2l^{2}%
}du\right)  P_{0}+duP_{1}+rd\phi P_{2}\label{A our gauge}\\
&  +\left(  \mathcal{W}d\phi+\frac{2}{l^{2}}\mathcal{Q}du\right)
J_{00}+\left(  \mathcal{W}du+2\mathcal{Q}d\phi\right)  P_{00}.\nonumber
\end{align}
Here, the arbitrary functions of $u$ and $\phi$ have been conveniently
redefined as%
\begin{align}
\mathcal{M}  & =2(\Xi_{+}+\Xi_{-}),\quad\mathcal{N}=l(\Xi_{+}-\Xi
_{-}),\label{MN}\\
\mathcal{W}  & =2(W_{+}+W_{-}),\quad\mathcal{Q}=l(W_{+}-W_{-}),\label{WQ}%
\end{align}
and then, the chirality conditions now read
\begin{align}
&  \partial_{u}\mathcal{M}=\frac{2}{l^{2}}\partial_{\phi}\mathcal{N}%
\quad2\partial_{u}\mathcal{N}=\partial_{\phi}\mathcal{M},\label{chirality 1}\\
&  \partial_{u}\mathcal{W}=\frac{2}{l^{2}}\partial_{\phi}\mathcal{Q}%
\quad2\partial_{u}\mathcal{Q}=\partial_{\phi}\mathcal{W}.\label{chirality 2}%
\end{align}
From eq. (\ref{A our gauge}) the generalized dreibein can be directly read, so
that the asymptotic form of the metric and the spin-3 field are recovered from
eqs. (\ref{metric}), (\ref{spin3 field}), which leads to%
\begin{align*}
ds^{2}  & =\left(  -\frac{r^{2}}{l^{2}}+\mathcal{M}\right)  du^{2}%
-2dudr+2\mathcal{N}dud\phi+r^{2}d\phi^{2},\\
\varphi & =(\mathcal{W}du+2\mathcal{Q}d\phi)du^{2}.
\end{align*}
One then verifies that, unlike what occurs in the highest weight gauge, and as
in the previous section, for our gauge choice the fields decouple, in the
sense that the spin-3 field does not generate any back-reaction on the metric.
Moreover, since the canonical charges, written in appendix \ref{limit charge},
do not change under the gauge transformation generated by $g_{\pm}$ in eqs.
(\ref{g+}), (\ref{g-}), the spin-3 field gives no contribution to the energy
and angular momentum.

One of the main advantages of expressing the asymptotic form of the connection
in our gauge choice is that the vanishing cosmological constant limit can be
taken in a straightforward way. This can be seen as follows. In the limit
$l\rightarrow\infty$, the chirality conditions (\ref{chirality 1}),
(\ref{chirality 2}) imply that%
\begin{equation}
\mathcal{M}=\mathcal{M}(\phi),\quad\mathcal{N}=\mathcal{J}(\phi)+\frac{u}%
{2}\partial_{\phi}\mathcal{M},\quad\mathcal{W}=\mathcal{W}(\phi),\quad
\mathcal{Q}=\frac{1}{2}(\mathcal{V}(\phi)+u\partial_{\phi}\mathcal{W}%
).\label{MJWV}%
\end{equation}
Hence, replacing (\ref{MJWV}) into the connection (\ref{A our gauge}), it is
simple to verify that for $l\rightarrow\infty$ the asymptotically flat
conditions in eq. (\ref{arriba}), are recovered. This is also the case for the
global charges which, as explained in appendix \ref{limit charge}, after a
suitable mapping of the functions that parametrize the asymptotic symmetries,
reduce to (\ref{ss9}) in the limit.

It is also very simple to check that the asymptotic symmetries, described by
two copies of the $W_{3}$ algebra (\ref{W3}), reduce to the spin-3 extension
of BMS$_{3}$. Indeed, by redefining the generators as
\begin{align}
P_{n} &  =\frac{1}{l}(\mathcal{L}_{n}^{+}+\mathcal{L}_{-n}^{-}),\\
J_{n} &  =\mathcal{L}_{n}^{+}-\mathcal{L}_{-n}^{-},\\
W_{n} &  =\frac{1}{l}(\mathcal{W}_{n}^{+}+\mathcal{W}_{-n}^{-}),\\
V_{n} &  =\mathcal{W}_{n}^{+}-\mathcal{W}_{-n}^{-},
\end{align}
the asymptotic symmetry algebra in the flat case, given by eqs. (\ref{ss14}),
(\ref{ss16}), (\ref{ss17}) is recovered once we take the limit $l\rightarrow
\infty$.

\bigskip

An additional benefit of this procedure, is that the asymptotic behavior of
higher spin gravity with $\Lambda=0$, for spins $s\geq2$, can be readily
obtained from the connection in the asymptotically AdS$_{3}$ case with our
gauge choice, as in eq. (\ref{A AdS our gauge}). Indeed, following the steps
described above, asymptotically flat spacetimes in higher spin gravity with
vanishing cosmological constant are found to be described by the following connection%

\begin{align}
A=\left( \frac{1}{2}\mathcal{M}du-dr \right. &\left. +\left(  \mathcal{J}+\frac{u}{2}%
\partial_{\phi}\mathcal{M}\right)  d\phi \right) P_{0}+du P_{1}+rd\phi P_{2}+\frac
{1}{2}\mathcal{M}d\phi J_{0}+d\phi J_{1}\nonumber\\
& +\left(  \mathcal{W}_{3}du+\left(  \mathcal{V}_{3}+u\partial_{\phi
}\mathcal{W}_{3}\right)  d\phi\right)  P_{00}+\mathcal{W}_{3}d\phi
J_{00}\label{A flat reloaded}\\
& +\left(  \mathcal{W}_{4}du+\left(  \mathcal{V}_{4}+u\partial_{\phi
}\mathcal{W}_{4}\right)  d\phi\right)  P_{000}+\mathcal{W}_{4}d\phi
J_{000}\nonumber\\
& +...\nonumber
\end{align}
where the arbitrary functions $\mathcal{M}$, $\mathcal{J}$, $\mathcal{W}_{3}$,
$\mathcal{V}_{3}$, $\mathcal{W}_{4}$, $\mathcal{V}_{4}$, ..., depend only on
$\phi$, and the generators $P_{a_{1}\cdot\cdot\cdot a_{s-1}}$, $J_{a_{1}%
\cdot\cdot\cdot a_{s-1}}$, are fully symmetric and traceless. The asymptotic
symmetry algebra that corresponds to the higher spin extension of BMS$_{3}$ is
then obtained from the one from the asymptotically AdS$_{3}$ case, spanned by
two copies of W$_{N}$, or W$_{\infty}\left[  \lambda\right]  $, by redefining
the generators as
\begin{align}
P_{n} &  =\frac{1}{l}(\mathcal{L}_{n}^{+}+\mathcal{L}_{-n}^{-}),\\
J_{n} &  =\mathcal{L}_{n}^{+}-\mathcal{L}_{-n}^{-},\\
W_{n}^{(3)} &  =\frac{1}{l}(\mathcal{W}_{n}^{(3)+}+\mathcal{W}_{-n}^{(3)-}),\\
V_{n}^{(3)} &  =\mathcal{W}_{n}^{(3)+}-\mathcal{W}_{-n}^{(3)-},\\
W_{n}^{(4)} &  =\frac{1}{l}(\mathcal{W}_{n}^{(4)+}+\mathcal{W}_{-n}^{(4)-}),\\
V_{n}^{(4)} &  =\mathcal{W}_{n}^{(4)+}-\mathcal{W}_{-n}^{(4)-},\\
&  ...
\end{align}
and then taking the limit $l\rightarrow\infty$.

\section{Discussion}

\label{Discussion}

The asymptotically AdS$_{3}$ conditions of \cite{Henneaux-HS,
Theisen-HS,Campoleoni-HS,Gaberdiel-Hartman}, that extend the ones of Brown and
Henneaux to the case of higher spin fields, have been shown to possess an
interesting vanishing cosmological constant limit. As a consequence, this
means that solutions of the field equations with $\Lambda<0$ that fulfill
these conditions, can be consistently mapped to asymptotically flat ones that
satisfy ours. Then, in particular, it is worth pointing out that the smoothed
out conical defects and surpluses discussed in
\cite{Castro-Gopakumar,Campoleoni1, Campoleoni2} fall within this category,
i.e., they admit a consistent $l\rightarrow\infty$ limit that fulfills the
field equations of higher spin gravity around flat space with the boundary
conditions discussed here. It is then natural to expect that their holonomies
around spacelike cycles remain trivial after the limit, but nonetheless, an
explicit check of this claim would be worth to be done.

Note that, as it has been recently shown in \cite{Liouville-dual-plano}, the
fact that General Relativity possesses a set of asymptotic conditions
fulfilling the BMS$_{3}$ algebra, implies that the theory can be described in
terms of a flat analogue of Liouville theory at null infinity, that, as shown
in \cite{Liouville-bms}, corresponds to a suitable limit of its AdS$_{3}$
counterpart \cite{Cousaert-Henneaux-V}. This strongly suggests that a similar
construction could be performed starting from the asymptotically flat
conditions in presence of higher spin fields discussed here, which would
naturally be identified with a flat analogue of Toda theory.

As an ending remark, it would be worth exploring whether the asymptotic
conditions proposed here could be generalized in a consistent way with the
asymptotic symmetries. In this sense, we would like pointing out that, at
least in the case of spins $s=2$, $3$, a good starting point might be the
generic solution of the field equations that fulfills the conditions
\begin{equation}
\omega_{u}^{a}=0,\quad e_{u}^{a}=\omega_{\phi}^{a},\label{ss2}%
\end{equation}%
\begin{equation}
W_{u}^{ab}=0,\quad E_{u}^{ab}=W_{\phi}^{ab},
\end{equation}
which ensure that the action attains an extremum. The solution is given by
\begin{equation}%
\begin{split}
W^{ab} &  =\mathcal{W}^{ab}d\phi,\\
E^{00} &  =\mathcal{W}^{00}du+\left(  \mathcal{V}^{00}+u\partial{_{\phi}%
}\mathcal{W}^{00}-2r\mathcal{W}^{02}\right)  d\phi,\\
E^{01} &  =\mathcal{W}^{01}du+\left(  \mathcal{V}^{01}+u\partial{_{\phi}%
}\mathcal{W}^{01}-r\mathcal{W}^{12}\right)  d\phi,\\
E^{02} &  =\mathcal{W}^{02}du+\left(  \mathcal{V}^{02}+u\partial{_{\phi}%
}\mathcal{W}^{02}+3r\mathcal{W}^{01}\right)  d\phi,\\
E^{11} &  =\mathcal{W}^{11}du+\left(  \mathcal{V}^{11}+u\partial{_{\phi}%
}\mathcal{W}^{11}\right)  d\phi,\\
E^{12} &  =\mathcal{W}^{12}du+\left(  \mathcal{V}^{12}+u\partial{_{\phi}%
}\mathcal{W}^{12}+r\mathcal{W}^{11}\right)  d\phi,
\end{split}
\label{ss3}%
\end{equation}
where $\mathcal{W}^{ab}$, $\mathcal{V}^{ab}$ are arbitrary functions of $\phi
$, while $e^{a}$ and $\omega^{a}$ are given by the ones of General Relativity
with $\Lambda=0$.

\bigskip

\acknowledgments We thank G. Barnich, M. Ba\~{n}ados, E. Bergshoeff, R. Canto,
A. Campoleoni, A. Castro, C. Erices, J. Gamboa, M. Gary, D. Grumiller, M.
Henneaux, C. Mart\'{\i}nez, A. P\'{e}rez, P. Sundell, D. Tempo, and C.
Troessaert for useful discussions and comments. J.M. thanks Conicyt for
financial support. M.P. is partially funded by Conicyt, project 7912010045. 
 R.T. thanks the Galileo Galilei Institute for Theoretical
Physics, the Physique th\'{e}orique et math\'{e}matique group of the
Universit\'{e} Libre de Bruxelles, and the International Solvay Institutes for
the kind hospitality. This research has been partially supported by Fondecyt
grants N${^{\circ}}$ 1130658, 1121031. Centro de Estudios Cient\'{\i}ficos
(CECs) is funded by the Chilean Government through the Centers of Excellence
Base Financing Program of Conicyt.

\appendix

\section{Higher spin extension of the Poincar\'{e} algebra as a contraction of
$sl(3)\oplus sl(3)$}

\label{contraction}

The algebra (\ref{CS1}) can be easily seen to arise from a contraction of two
copies of $sl(3)$, whose generators fulfill%
\begin{align}
\lbrack J_{a}^{\pm},J_{b}^{\pm}]  & =\epsilon_{abc}J^{\pm c},\label{CS4}\\
{}[J_{a}^{\pm},T_{bc}^{\pm}]  & =\epsilon_{\;\;a(b}^{m}T_{c)m}^{\pm},\\
{}[T_{ab}^{\pm},T_{cd}^{\pm}]  & =-\left(  \eta_{a(c}\epsilon_{d)bm}%
+\eta_{b(c}\epsilon_{d)am}\right)  J^{\pm m}.
\end{align}
This can be performed by changing the basis of the $sl(3)\oplus sl(3)$
generators according to%
\begin{align}
& P_{a}=\frac{1}{l}(J_{a}^{+}-J_{a}^{-}),\quad J_{a}=J_{a}^{+}+J_{a}%
^{-},\label{CS5}\\
& P_{ab}=\frac{1}{l}(T_{ab}^{+}-T_{ab}^{-}),\quad J_{ab}=T_{ab}^{+}+T_{ab}%
^{-},
\end{align}
so that after taking the limit $l\rightarrow\infty$, the algebra (\ref{CS1})
is recovered.

\section{Lie-algebra-valued parameter components of the asymptotically flat
symmetries}

\label{appendix lambda}

The components of the Lie-algebra-valued parameter $\lambda=\lambda
(\epsilon,y,w,v)$ in eq. (\ref{lambda-flat}), associated to the asymptotically
flat symmetries, depend on four arbitrary functions of $\phi$ and its
derivatives. They are given by
\begin{align}
\rho^{0} &  =-y^{\prime\prime\prime}u+(r+\frac{u}{2}\mathcal{M})y^{\prime
}+4vu\mathcal{W}^{\prime}+\frac{1}{2}yu\mathcal{M}^{\prime}+4u \mathcal{W}%
v^{\prime}-\epsilon^{\prime\prime}+y\mathcal{J}\nonumber\\
&  \;\;\;\;+\frac{1}{2}\mathcal{M}\epsilon+4 v\mathcal{V}+4 \mathcal{W}w,\\
\rho^{1} &  =\epsilon+uy^{\prime},\\
\rho^{2} &  =ry-y^{\prime\prime}u-\epsilon^{\prime},\\
\eta^{0} &  =-y^{\prime\prime}+\frac{1}{2}\mathcal{M}y+4 \mathcal{W}v,\\
\eta^{1} &  =y,\\
\eta^{2} &  =-y^{\prime},\\
\xi^{00} &  =\frac{1}{6}uv^{(5)}-\frac{1}{3}(r+2u\mathcal{M})v^{\prime
\prime\prime}-\frac{1}{6}u\mathcal{M}^{\prime\prime\prime}v-\frac{3}%
{4}u\mathcal{M}^{\prime\prime}v^{\prime}-\frac{5}{4}u\mathcal{M}^{\prime
}v^{\prime\prime}+v(\frac{r}{3}+\frac{u}{2}\mathcal{M})\mathcal{M}^{\prime
}\nonumber\\
&  \;\;\;\;+\frac{1}{12}(10r\mathcal{M}+3\mathcal{M}^{2}u)v^{\prime
}+u\mathcal{W}y^{\prime}+\mathcal{W}\epsilon+y\mathcal{V}-\frac{4}%
{3}\mathcal{J}v^{\prime\prime}+J\mathcal{M}v-\frac{1}{3}v\mathcal{J}%
^{\prime\prime}-\frac{1}{6}\mathcal{M}^{\prime\prime}w\nonumber\\
&  \;\;\;\;-{\frac{7}{12}}\mathcal{M}^{\prime}w^{\prime}-\frac{2}%
{3}\mathcal{M}w^{\prime\prime}-\frac{7}{6}v^{\prime}\mathcal{J}^{\prime}%
+\frac{1}{4}\mathcal{M}^{2}w+\frac{1}{6}w^{(4)}+u\mathcal{W}^{\prime}y,\\
\xi^{01} &  =\frac{1}{2}rv^{\prime}-\frac{1}{6}v^{\prime\prime\prime}%
u-\frac{1}{6}w^{\prime\prime}+\frac{1}{6}\mathcal{M}v^{\prime}u+\frac{1}%
{6}\mathcal{M}w+\frac{1}{3}v\mathcal{J}+\frac{1}{6}vu\mathcal{M}^{\prime},\\
\xi^{02} &  =-\frac{1}{2}rv^{\prime\prime}+\frac{1}{6}v^{(4)}u+\frac{1}%
{6}w^{\prime\prime\prime}-{\frac{7}{12}}v^{\prime}u\mathcal{M}^{\prime}%
-{\frac{5}{12}}\mathcal{M}v^{\prime\prime}u-\frac{1}{6}\mathcal{M}^{\prime
}w-{\frac{5}{12}}\mathcal{M}w^{\prime}-\frac{5}{6}v^{\prime}\mathcal{J}%
\nonumber\\
&  \;\;\;\;-\frac{1}{3}v\mathcal{J}^{\prime}-\frac{1}{6}vu\mathcal{M}%
^{\prime\prime}+\frac{1}{2}\mathcal{M}rv,\\
\xi^{11} &  =w+uv^{\prime},\\
\xi^{12} &  =rv-\frac{1}{2}v^{\prime\prime}u-\frac{1}{2}w^{\prime},\\
\Lambda^{00} &  =\frac{1}{6}v^{(4)}-\frac{1}{6}\mathcal{M}^{\prime\prime
}v-{\frac{7}{12}}\mathcal{M}^{\prime}v^{\prime}-\frac{2}{3}\mathcal{M}%
v^{\prime\prime}+\frac{1}{4}\mathcal{M}^{2}v+\mathcal{W}y,\\
\Lambda^{01} &  =-\frac{1}{6}v^{\prime\prime}+\frac{1}{6}\mathcal{M}v,\\
\Lambda^{02} &  =\frac{1}{6}v^{\prime\prime\prime}-\frac{1}{6}\mathcal{M}%
^{\prime}v-{\frac{5}{12}}\mathcal{M}v^{\prime},\\
\Lambda^{11} &  =v,\\
\Lambda^{12} &  =-\frac{1}{2}v^{\prime}.
\end{align}

\section{Global charges and its vanishing cosmological constant limit}

\label{limit charge}

In the case of the asymptotically AdS$_{3}$ conditions proposed in
\cite{Henneaux-HS,Theisen-HS}, the expression for the global charges is given
by%
\[
Q=\frac{kl}{2\pi}\int d\phi\left[  \varepsilon^{+}\Xi_{+}+\varepsilon^{-}%
\Xi_{-}+4(\chi^{+}W_{+}+\chi^{-}W_{-})\right]  ,
\]
which, as expected, does not change for the gauge choice described in section
\ref{Limit}. Here, the chiral functions $\varepsilon^{\pm}(x^{\pm})$ and
$\chi^{\pm}(x^{\pm})$ correspond to the asymptotic symmetries spanned by the
components of the Lie-algebra-valued parameter along the generators $L_{\pm
1}^{\pm}$ and $W_{\pm2}^{\pm}$,  respectively. In order to take the limit
$l\rightarrow\infty$, it is convenient to express the charges according to the
definitions in eqs. (\ref{MN}), (\ref{WQ}), i.e.,%
\begin{equation}
Q=\frac{k}{4\pi}\int d\phi\left[  f\mathcal{M}+2y\mathcal{N}+4(h\mathcal{W}%
+2v\mathcal{Q})\right]  ,\label{cargasADS}%
\end{equation}
so that the functions that parametrize the asymptotic symmetries become
naturally redefined as%
\[
f=\frac{l}{2}(\varepsilon^{+}+\varepsilon^{-}),\quad y=\frac{1}{2}%
(\varepsilon^{+}-\varepsilon^{-}),\quad h=\frac{l}{2}(\chi^{+}+\chi^{-}),\quad
v=\frac{1}{2}(\chi^{+}-\chi^{-}),
\]
and hence, the chirality conditions now read%
\[
\partial_{u}f=\partial_{\phi}y,\quad\partial_{u}y=\frac{1}{l^{2}}%
\partial_{\phi}f,\quad\partial_{u}h=\partial_{\phi}v,\quad\partial_{u}%
v=\frac{1}{l^{2}}\partial_{\phi}h\quad.
\]
Note that for $l\rightarrow\infty$, the latter conditions imply that%
\[
y=y(\phi),\quad f=\epsilon(\phi)+uy^{\prime},\quad v=v(\phi),\quad
h=w(\phi)+uv^{\prime},
\]
and therefore, by virtue of the corresponding relationship of the functions
$\mathcal{M}$, $\mathcal{N}$, $\mathcal{W}$ and $\mathcal{Q}$, in eq.
(\ref{MJWV}), the global charges (\ref{cargasADS}) reduce to ours in eq.
(\ref{ss9}) in the vanishing cosmological constant limit.

\end{document}